\documentclass[twocolumn,showpacs,amsmath,amssymb,pre,floatfix,pre,superscriptaddress]{revtex4-1}

\usepackage{color}
\usepackage[normalem]{ulem}
\usepackage{graphicx}
\begin{document}
\title{Polydispersed rods on the square lattice}    

\author{J\"urgen F. Stilck}
\email{jstilck@if.uff.br}
\affiliation{Instituto de F\'{\i}sica and
National Institute of Science and
Technology for Complex Systems,
Universidade Federal Fluminense,
Av. Litor\^anea s/n,
24210-346 - Niter\'oi, RJ,
Brazil}
\author{R. Rajesh}
\email{rrajesh@imsc.res.in}
\affiliation{The Institute of Mathematical Sciences, 
C.I.T. Campus, Taramani, Chennai 600113, India}

\date{\today}

\begin{abstract}
We study the grand-canonical solution of a system of hard polydispersed 
rods placed on the square lattice using transfer matrix and finite size 
scaling calculations. We determine the critical line 
separating an isotropic 
from a nematic phase. No second transition to a disordered phase is
found at high density, contrary to what is observed in  
the monodispersed case. The 
estimates of critical exponents and the central charge on the critical 
line are consistent with the Ising universality class.
\end{abstract}


\maketitle

\section{\label{sec:intro}Introduction and definition of the model}

The question of the existence of an ordering transition in a system of 
long rigid 
rods with only excluded volume interactions is
relevant to the study of many physical systems. Examples include  
liquid crystals~\cite{degennesBook}, tobacco
mosaic
virus~\cite{fraden1989} and carbon
nanotube gels~\cite{islam2004}. In a seminal paper, Onsager 
showed that,
for large enough 
density and aspect ratio of cylindrical rods, the system undergoes a 
discontinuous phase transition from an
isotropic to an orientationally ordered nematic phase in 
three dimensions~\cite{o49}. The case of discrete positions of the rods was 
considered by Flory~\cite{f56} and a model of rods in a continuum space 
but with a discrete set of orientations was studied by Zwanzig~\cite{z63}, 
with similar results. A review of these continuum models may 
be found in Ref.~\cite{vl92}. In two dimensions, with continuous 
positions and orientations of the rods, it is known that the rotational 
symmetry can not be broken~\cite{mw66}, but a 
Berezinskii-Kosterliz-Thouless transition~\cite{b71,b72,kt73} is found 
between a low density phase with exponentially decaying orientational 
correlation and a high density phase where the correlations decay with a 
power law~\cite{straley1971,frenkel1985,khandkar2005,vink2009}.

The lattice version of the problem, where the rods are composed of $k$ 
collinear and consecutive sites of a regular lattice ($k$-mers), has 
also been studied in the literature. Here, a lattice
site may be occupied by utmost one $k$-mer.
For the particular case of dimers 
($k=2$), it is known that there is no ordering transition at finite
density of unoccupied sites
in any dimension~\cite{Heilmann1970,Kunz1970,Gruber1971,lieb1972},
even though additional interactions can result in an
ordering transition~\cite{d12,jesper2005,fradkin2014}. 
For longer rods, Ghosh and Dhar~\cite{gd07}
numerically showed the existence of a nematic phase for $k \ge
k_{\mbox{min}}$ where $k_{\mbox{min}}=7$ on a square lattice,
following which
the existence of the nematic phase has been proved 
rigorously for $k \gg 1$~\cite{dg13}.
However,
the fully packed phase is expected to be
disordered~\cite{degennesBook,gd07}, resulting in a second transition
from nematic to disordered phase  at high densities. This has been
demonstrated numerically~\cite{krds12,krds13}.
A solution of the model on a Bethe-like lattice leads to 
continuous or discontinuous isotropic-nematic transitions for 
sufficiently high values of $k$, depending of the coordination number of 
the lattice. The second
transition does not occur on such a 
lattice~\cite{drs11}, although two transitions are found on a Bethe-like 
lattice if additional repulsive interactions between the rods are 
included~\cite{kr13}. 

Detailed Monte Carlo studies of the first transition 
show that it is in the Ising universality class on the square lattice
and the three state Potts universality class on the triangular and hexagonal
lattices, as
would be expected from the symmetry of the ordered
phase~\cite{flr08,flr08a,fernandez2008c,linares2008,fischer2009}.
The 
study of the second transition 
using simulations is more difficult 
due to the presence of
long-lived metastable
states, but these difficulties were overcome using an 
efficient grand-canonical Monte Carlo algorithm~\cite{krds12,krds13}. 
On the triangular lattice these simulations indicate that the 
high density transition is also in the three-states Potts universality 
class, but on the square lattice, due to strong orientational 
correlations in the high density disordered phase, although non-Ising 
effective exponents were found, it was not possible to rule out  a 
crossover to Ising exponents  at larger system sizes  
than what were studied.
Another generalization of the 
model is to consider hard rectangles, comprising $m \times mk$ sites on 
the square lattice~\cite{kr14,kr14b,nkr14}, and in some cases 
up to three entropy driven
transitions were found with increasing density for $m>1$.

In all the cases discussed above,
the rods were monodispersed, but 
polydispersity in rod lengths 
may hardly
be avoided in experimental 
situations~\cite{ba93,bkl96}. 
From a theoretical point of view its effects on the phase 
transitions found in the models are interesting~\cite{vl92}. 
In continuous models,
calculations show that polydispersity
may have strong effects on the 
behavior of the system, such as a larger density gap of the 
discontinuous nematic transition, higher nematic order parameter for the 
longer rods in the nematic phase, reentrant nematic phase and even two 
distinct nematic phases coexisting with the isotropic 
phase~\cite{l84,Birshtein1988,sollich2003_01,sollich2003_02}.  
Similar
results are obtained within the polydispersed Zwanzig model with restricted
orientations~\cite{sear2000,cuesta2003}.

Polydispersed systems on lattices are less well studied. One 
question of interest is whether a second transition
to the high density disordered phase, present in
monodispersed systems, will persist in a polydispersed
system. We recall the argument for
the fully packed limit of monodispersed systems being 
disordered~\cite{degennesBook,gd07}. In its simplest version, 
it consists of dividing
the lattice into $k \times k$ squares, so 
that in each square the $k$ rods may be all either 
horizontal or vertical.  This establishes a lower bound to the entropy per site 
of the disordered 
phase $s \ge k^{-2}  \ln 2$, which is larger than the vanishing entropy of 
the fully ordered nematic phase. For polydispersed system, the 
fully nematic phase clearly has non-zero entropy, making it 
unclear its effect on
the high density transition. If the high density phase persists, then it raises
the interesting scenario of the fully packed system undergoing a 
isotropic--nematic transition.

In this paper, we study the phase transitions and critical behavior of a
particular  model of polydispersed hard rods on the square lattice.
The model we consider is inspired by the equilibrium polymerization 
model that was used, among other applications,
to study the polymerization transition of liquid sulfur~\cite{wkp80,wp81}. The 
statistical weight of a rod comprising $k$ successive collinear sites is 
equal to $z_i^{k-2} z_e^2$, where $z_i$ is the activity of an internal 
monomer and $z_e$ the activity of an monomer at an endpoint of a rod. 
No two rods may overlap. The limit  $z_i=0$ 
maps onto the dimer model, while the limit $z_e \to 0$ corresponds to rods
of infinite length.
In general, the mean
length of rods and 
the density of occupied lattice sites will be determined by the 
activities $z_i$ and $z_e$. We obtain the   
thermodynamic behavior of the system by  finding the solution on 
strips of width $L$ 
using a transfer matrix method and then 
applying
finite size scaling to extrapolate the results to the limit of infinite width.
We 
show that the system undergoes a continuous isotropic-nematic phase 
transition as the density of occupied sites is increased, provided the 
mean number of monomers per rod is high enough. 
Estimates
of the critical exponents are consistent with the transition being in the
the Ising universality class. No high density 
transition from the nematic phase to an isotropic phase is found
within this model.

An exact solution exists for a particular case of the model, when it
may be mapped onto the two dimensional Ising model~\cite{i06}.
In this solvable case, the 
activity of a rod with $k$ monomers is $2^{-k}q^{1-k}$. This corresponds to
the choice 
$z_e=\sqrt{z_i/2}$. A continuous Ising transition 
between an isotropic and a nematic phase occurs at a critical value 
$q_c=1/(2+2\sqrt{2})$, corresponding to  
$z_i=1+\sqrt{2}$, $z_e=\sqrt{(1+\sqrt{2})/2}$ being a point on the critical 
line~\cite{i06}.
Our transfer matrix results for this value of $z_i$ and $z_e$ are consistent 
with the exact solution.

The rest of the paper is organized as follows.
In Sec.~\ref{tm} we define the 
transfer matrix for the model on strips. In Sec.~\ref{cb} the critical
behavior of the model is obtained. The results for fully packed
limit is contained in Sec.~\ref{sec:full-packing} and those for finite
density in Sec.~\ref{sec:finite-density}.
Section~ \ref{conc} contains a discussion
of results and some open problems.

\section{Transfer matrix solution on strips}
\label{tm}

\begin{figure}
\centering
\includegraphics[width=8cm]{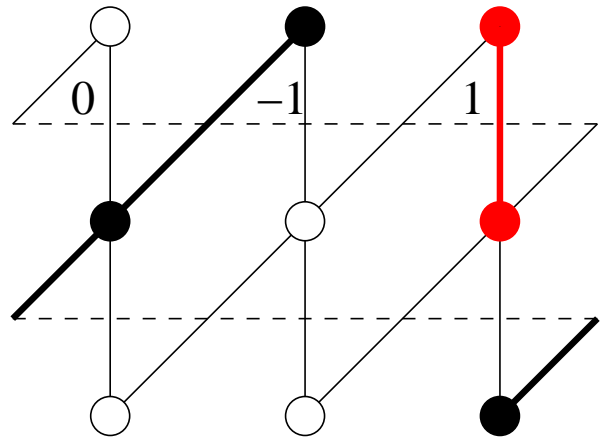}
\caption{A strip of width $L=3$ with periodic boundary
condition in the horizontal direction. The successive states are defined by the 
configurations of the edges crossing the dashed lines. 
Part of a rod in the $x$ direction (black) 
and of another in the $y$ direction (red) are shown. The indices 
$0,-1,1$ define the state of the upper set of 
$L$ pairs of $x$ and $y$ edges. The transfer is
in the vertical direction.
}
\label{f1}
\end{figure}

We study the model on strips of finite width $L$, and extrapolate 
our results to the two-dimensional thermodynamic
limit $L \to \infty$. We define the 
width of the strip as equal to the number of sites aligned horizontally 
across the strip. In order to treat both directions of the
square lattice symmetrically, we rotate the 
lattice by  $\pi/4$, so that 
after a convenient deformation the strip is
as shown in 
Fig.~\ref{f1}. The edges of the square lattice 
in the $y$-direction correspond
to vertical edges in Fig.~\ref{f1}, while the edges in $x$-direction are at
an angle $\pi/4$ with the horizontal in Fig.~\ref{f1}.
Periodic boundary conditions
are applied in the horizontal and vertical directions.
We note that, with respect to the original 
lattice, the transfer is in the diagonal direction.

The states used to build the transfer matrix are given by the possible 
configurations of the edges crossing a horizontal line between 
sets of horizontal sites, 
such as the two dashed lines depicted in Fig.~\ref{f1}. They are defined by 
the occupancy of the edges by bonds belonging to rods. We associate an 
integer number to each pair of $x$ and $y$ edges crossing the dashed 
line which meet at a lattice site above. Either both edges are empty 
(index $0$), the $x$ edge is occupied (index -1), or the $y$ edge is 
occupied (index $1$). For example, in the upper dashed line, no bond 
belonging to a rod is present in the first pair of lattice edges, from 
left to right. Thus, we associate the index 0 to this pair. In the 
second pair of edges an $x$-mer crosses the line, so that a index -1 is 
associated to this pair. Finally, a $y$-mer crosses the third pair, so 
that a index 1 is associated to it. Therefore, we denote the state 
of the upper dashed line as $(0,-1,1)$. Similarly,
the state of the lower dashed line is
$(-1,0,0)$. We note that 
all combinations of indices do not represent allowed 
states, for example, an index 1 
may not be followed by an index -1. 
When $L=3$, there are 18 different states.

The transfer matrix corresponds to the operation of adding a new row of 
$L$ sites to the lattice. The configuration of the row of sites between 
two successive states, as the ones defined by the dashed lines in 
Fig.~\ref{f1}, determine the matrix element. In the example of Fig.~\ref{f1}, there
is one internal monomer and one endpoint monomer, so that the matrix 
element associated to this pair of states is equal to $z_i z_e$.  In 
general, the transfer matrix for this model is not symmetric. 

Knowing the transfer matrix, the thermodynamic quantities may be
determined. The 
grand-canonical free energy per site (divided by $k_BT$) of the model 
in the thermodynamic limit is given by
\begin{equation}
\phi_L(z_i,z_e)=\lim_{L\to \infty} -\frac{1}{L} \ln \lambda_1(z_i,z_e),
\label{eq:phidefn}
\end{equation}
where $\lambda_1$ is the largest eigenvalue of the transfer matrix. The 
correlation length is given by:
\begin{equation}
\xi_L(z_i,z_e)=\frac{1}{\ln\left(\frac{\lambda_1}{|\lambda_2|}\right)},
\label{cl}
\end{equation}
where $\lambda_2$ is the eigenvalue with second largest
modulus.
In order to look for phase transitions in the two-dimensional limit $L 
\to \infty$, we use the phenomenological renormalization group argument 
and, for a given value of $z_e$, search for a value of $z_i$ where the 
fixed point relation
\begin{equation}
\frac{\xi_L}{L}=\frac{\xi_{L+1}}{L+1},
\label{prg}
\end{equation}
holds with $\xi$ as in Eq.~(\ref{cl}). 

From the largest eigenvalue, other relevant quantities may be 
calculated as follows. The densities of internal and endpoint monomers in 
a strip of width $L$ 
may be calculated by differentiating the leading eigenvalue with respect to the 
activities:
\begin{equation}
\rho_{i,L}=\frac{z_i}{L\lambda_{1,L}}\frac{\partial \lambda_{1,L}} 
{\partial z_i},
\end{equation}
\begin{equation}
\rho_{e,L}=\frac{z_e}{L\lambda_{1,L}}\frac{\partial \lambda_{1,L}} 
{\partial z_e}.
\end{equation}
The mean number of monomers per rod is given by:
\begin{equation}
\langle k \rangle=\frac{2 \rho}{\rho_e},
\end{equation}
where $\rho=\rho_i+\rho_e$ is the fraction of lattice sites occupied by 
monomers.

The limit of full packing is reached when both activities 
diverge. More precisely, we consider the limit $z_e,z_i \to \infty$ with 
a fixed ratio 
\begin{equation}
r=\frac{z_e}{z_i},~z_e, z_i \to \infty.
\label{eq:rdefn}
\end{equation}
This limit may be studied using transfer 
matrices by taking into account only contributions where all sites are 
occupied by monomers. From the largest eigenvalue, the fraction of 
endpoint monomers in the fully packed limit may be found using
\begin{equation}
\rho_{e,L}=\frac{r}{L\lambda_{1,L}}\frac{\partial \lambda_{1,L}}{\partial r}, ~\rho=1.
\end{equation}
The mean number of monomers per rod in this limit is:
\begin{equation}
\langle k \rangle =\frac{2}{\rho_e},~\rho=1.
\label{kmed-fl}
\end{equation}

In the limit of rods of infinite length ($z_e \to 0$), 
all eigenvalues may be found easily. 
It is straightforward to see that the only non-vanishing
entries of the transfer matrix are between states where all rods are in the
same direction. The total number of  states is then
$2^{L+1}-1$.

In the spectrum of this transfer matrix, there is one unitary 
eigenvalue 
and
\[
2\frac{L!}{m!(L-m)!}
\]
eigenvalues of modulus $z_i^m$, with $m=1,2,\ldots,L$. The remaining 
eigenvalues equal zero. Therefore, in this limit, 
a discontinuous phase 
transition occurs at $z_i=1$ between the empty 
lattice and a 
saturated nematic phase where the whole lattice is filled with $L$ rods 
in one of the two directions. When $z_e >0$, we do not find any 
value of $z_i$ for which the leading eigenvalue is degenerate, although 
there are vanishing elements in the transfer matrix and thus the 
hypothesis of the Perron-Frobenius theorem is not fulfilled.

The elements of the transfer matrix are monomials in the 
activities $z_e$ and $z_i$, and are determined exactly for given widths 
$L$ using a computer code. In Table~\ref{dimen} we show the numbers 
of states and the numbers of lines of the file with integers defining 
the transfer matrix, for $L=3,4,\ldots,9$ in the general case and for 
$L=3,4,\ldots,12$ if only fully packed  configurations are allowed. The 
fraction of non-vanishing elements of the matrix in the general and full 
lattice cases are also shown in Table~\ref{dimen}, and it is clear that the 
matrices are quite sparse. This favors the use of algorithms related to 
the power method to find the leading eigenvalues of the transfer matrix.
\begin{table}
\caption{\label{dimen} The numbers of states ($N_s$) and numbers of 
lines ($N_l$) in the files defining the transfer matrices for the values 
of the widths ($L$) we considered in the phenomenological
renormalization group calculations. $N_l-N_s$ is
the number of non-zero entries in the transfer matrix. The fraction 
$f$ of non-zero elements of the matrix is also given. In the last 
two columns, we indicate the number of lines of the transfer matrix and 
the fraction of non-vanishing elements restricted to full lattice 
configurations.}
\begin{ruledtabular}
\begin{tabular}{|c|c|c|c|c|c|}
$L$&$N_s$&$N_l$&$f$&$N_{l,fl}$&$f_{fl}$\\
\hline
3&18&217&0.6142&114&0.2963\\
4&47&1202&0.5229&481&0.1965\\
5&123&6850&0.4446&2099&0.1306\\
6&322&39525&0.3781&9332&0.08690\\
7&843&229330&0.3215&41939&0.05783\\
8&2207&1333922&0.2734&189665&0.03849\\
9&5778&7767577&0.2325&860874&0.02561\\
10&15127&45254202&0.1977&3915689&0.01705\\
11&39603&-&-&17832219&0.01124\\
12&103682&-&-&81265636&0.007550\\
\end{tabular}
\end{ruledtabular}
\end{table}

In order to determine numerically the two leading eigenvalues of the 
transfer matrix, we use a modified version of the algorithm due to 
Gubenatis and Booth~\cite{gb08}. A brief description of the algorithm
is presented in the Appendix. As in all procedures based on the power 
method, the convergence rate is determined by the spacing between the 
leading eigenvalues, so that the rates are very low close to the 
discontinuous transition at $z_e \to 0$. 

\section{\label{cb} Critical behavior, extrapolation of critical line, 
and densities}

In this section, we determine the critical values of the parameters
and the exponents describing the 
transition from transfer matrix calculations on finite strips.
The results for the fully packed limit are in
Sec.~\ref{sec:full-packing} and those for other densities in
Sec.~\ref{sec:finite-density}.

\subsection{\label{sec:full-packing} Full Packing}

\begin{figure}
\centering
\includegraphics[width=8cm]{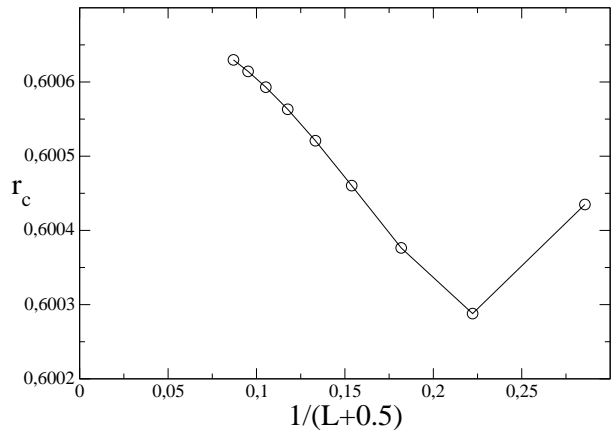}
\caption{Estimates for the critical value of the ratio of activities $r$ 
in the full lattice limit.}
\label{rcrit-fl}
\end{figure}

We first estimate the critical value for the ratio of the activities $r_c$
[see Eq.~(\ref{eq:rdefn})] in the fully packed limit.
Figure~\ref{rcrit-fl} shows estimates 
of $r_c$ as a function of $(L+0.5)^{-1}$, using pairs 
of widths $L$ and $L+1$, corresponding to strips of widths 
$L=3,4,\ldots,12$. A non-monotonic behavior is observed at small widths.
Therefore, we discard the results obtained from the pairs $L=3,4$ and $L=4,5$, 
restricting ourselves to the concave portion of the curve 
in Fig.~\ref{rcrit-fl}. 

\begin{figure}
\centering
\includegraphics[width=8cm]{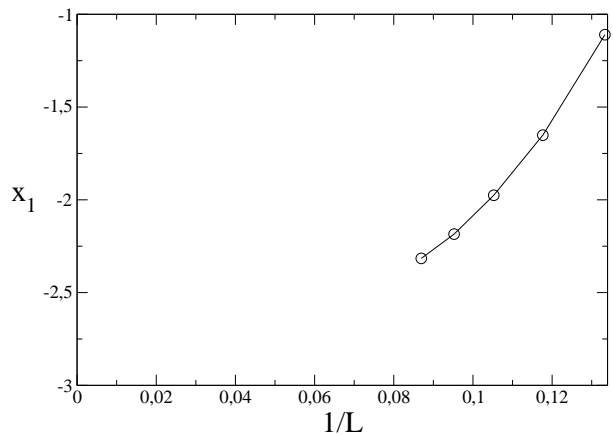}
\caption{Estimate of the exponent $x_1$, where  the critical
ratio $r_c(L)$ is fitted to the form $r_c(L)=r_c+aL^{x_1}$. The data
are for the fully packed limit.}
\label{extrap-fl}
\end{figure}

The data for $r_c(L)$ is fitted to the form
$r_c(L)=r_c+aL^{x_1}$ using three successive values of $L$. The estimates of
the parameters $r_c$, $a$ and $x_1$ from the three pairs 
$L,L+1$, $L+1,L+2$, and $L+2,L+3$ are 
associated to the width $L+3/2$.
The results for the 
exponent $x_1$ are shown  in Fig.~\ref{extrap-fl}. 
It is apparent that the data are still far from the 
asymptotic value. However, the value $x_1=-3$ is not ruled out by them, 
and considering the evidence presented later for the phase transition 
being in the Ising universality class, we assume $x_1=-3$~\cite{gb02}. 

\begin{figure}
\centering
\includegraphics[width=8cm]{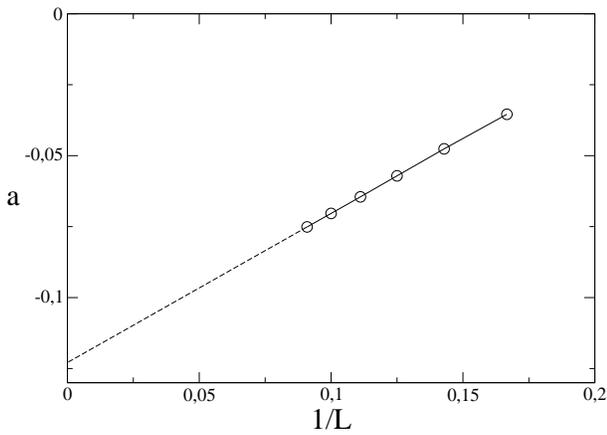}
\caption{Estimate of the amplitude $a$, where  the critical
ratio $r_c(L)$ is fitted to the form $r_c(L)=r_c+aL^{x_1}$.
The data are for the fully packed limit. The dashed line corresponds to
the linear extrapolation.
}
\label{extrap2p-fl}
\end{figure}

We then may obtain two point estimates by fitting the data 
to the function $r_c(L)=r_c+aL^{-3}$. The results for the amplitude $a$, 
presented in Fig.~\ref{extrap2p-fl}, show a clear linear behavior with $L^{-1}$.
We extrapolate the estimates to $L \to \infty$ to obtain 
$a=-0.1229$. Using this value for the amplitude in the behavior of 
the estimates for the critical ratio as a function of the width, we 
obtain an extrapolated value $r_c=0.6007$. We note that an unique 
solution is found for the fixed point of the recursion relation 
Eq.~(\ref{prg}), so that there are
no evidences for a second
transition at high $r$.

\begin{figure}
\centering
\includegraphics[width=8cm]{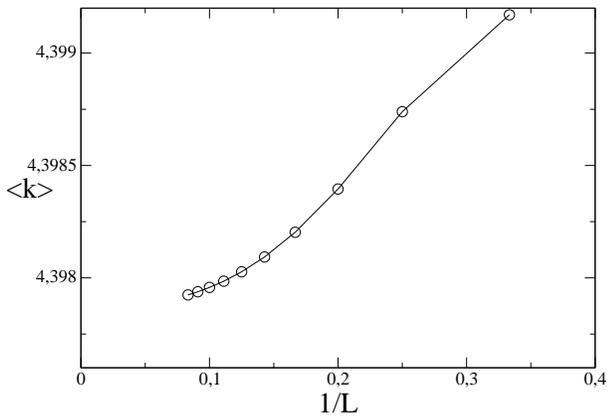}
\caption{Estimates for the critical value of the mean number of monomers 
per rod $\langle k \rangle$ in the limit of full packing. Results are for 
widths ranging from $L=3$ to $L=12$.}
\label{densi-fl}
\end{figure}

We now estimate the mean number of monomers per rod $\langle k
\rangle$ using Eq.~(\ref{kmed-fl})
at the estimated value of the critical ratio $r_c$. In 
Fig.~\ref{densi-fl}, its  variation  
with $L^{-1}$ is shown. The 
extrapolation to the two-dimensional limit, using three point fits to 
the data, gives $\langle k \rangle=4.3978$ at $r =r_c$.

The dimensionless free energy per site $\phi$ of the system 
in a strip of width  $L$ is given by Eq.~(\ref{eq:phidefn}).
Conformal invariance theory states that at the critical condition, for 
periodic boundary conditions, the free energy obeys~\cite{bcn86}:
\begin{equation}
\phi(L) \approx f - \frac{\pi c}{6 \ell^{2}}.
\label{eq:energy}
\end{equation}
where $c$ is the central charge, $f$ is the free energy per site in the 
two-dimensional limit $\ell \to \infty$, and $\ell$ is the width of the 
strip measured in units of the lattice parameter. In the usual 
orientation of the square lattice,  $\ell=L$, but in the present 
case, with the lattice rotated by an angle $\pi/4$, the width of a strip 
with $L$ rows of $L$ sites is $\ell=\sqrt{2} L$, so that we may rewrite 
the Eq.~(\ref{eq:energy}) as
\begin{equation}
\phi(L) \approx f - \frac{\pi c}{12 L^2}.
\end{equation}

\begin{figure}
\centering
\includegraphics[width=8cm]{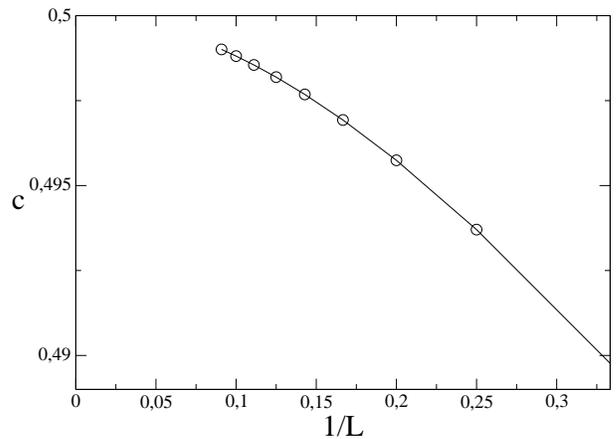}
\caption{Estimates of the central charge $c$ of the model in the limit of
full packing obtained from the free energies per site at the estimated 
critical ratio $r_c$.}
\label{c-fl-i}
\end{figure}

Estimates for the central charge may then be obtained by considering free 
energies at the extrapolated critical ratio for different widths. 
From the critical free energies for two widths, an estimate for 
$c$ may be found through:
\begin{equation}
c(L+1/2)=\frac{12}{\pi}\frac{\phi(L+1)-\phi(L)}{1/(L+1)^2-1/L^2}
\label{ce}
\end{equation}
The result is shown in Fig.~\ref{c-fl-i}, and the convergence of the 
data to the $c=1/2$, which corresponds to the Ising universality class, 
is apparent.

\begin{figure}
\centering
\includegraphics[width=8cm]{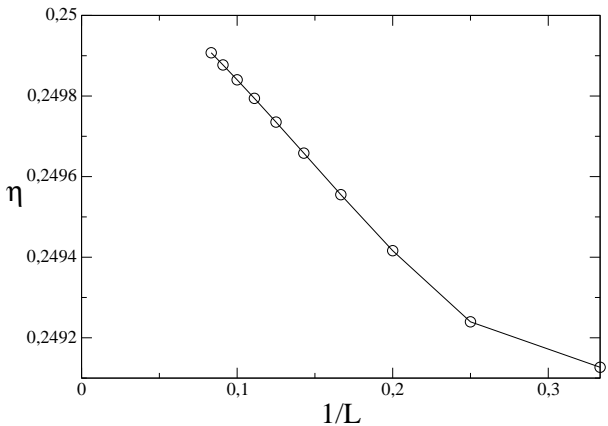}
\caption{Estimates of the critical exponent $\eta$ at full packing, 
obtained from the amplitudes of the finite size behavior of the 
correlation length at the estimated critical ratio $r_c$.}
\label{eta-fl}
\end{figure}

Another universal quantity which is rather simple to estimate is the 
amplitude of the finite size behavior of the inverse correlation length 
at criticality~\cite{nb83,c84}. The asymptotic behavior of the inverse 
correlation length is
\begin{equation}
\frac{1}{\xi_\ell} \sim \frac{A}{\ell},
\label{eta1}
\end{equation}
where the correlation length may be calculated using Eq.~(\ref{cl}). The 
amplitude $A$ is related to the critical exponent $\eta$:
\begin{equation}
A=\pi \eta.
\end{equation}
However, in this result it is implicit that both the correlation length 
and the width of the strip are measured in units of the lattice spacing. 
Since each time we apply the transfer matrix we increase the length of 
the strip by $\sqrt{2}/2$ lattice spacings, we have 
$\xi_\ell=\frac{\sqrt{2}}{2}\xi_L$ and, as discussed above 
$\ell=\sqrt{2}L$, so that we may rewrite Eq.~(\ref{eta1}) as:
\begin{equation}
\frac{1}{\xi_L} \sim \frac{A}{2L}.
\label{eta2}
\end{equation}
The estimate for the critical exponent $\eta$ are shown in 
Fig.~\ref{eta-fl}, as a function of $L^{-1}$. They are compatible with the value 
for the Ising universality class $\eta=1/4$.

\begin{figure}
\centering
\includegraphics[width=8cm]{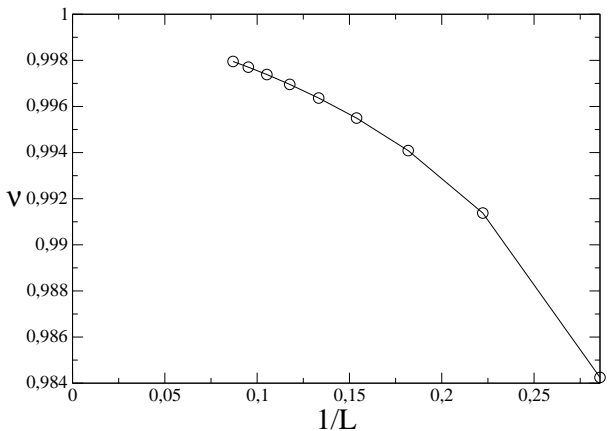}
\caption{Estimates of the critical exponent $\nu$ at full packing, obtained 
from Eq.~(\ref{nuest}), at the estimated 
critical ratio $r_c$. }
\label{nu-fl}
\end{figure}

Finally, we may also estimate the critical exponent $\nu$ which 
characterizes the divergence of the correlation length. This may be done 
by expanding the recursion relation of the phenomenological 
renormalization group close to the fixed point Eq.~(\ref{prg}):
\begin{equation}
\nu_L \approx
 \frac{\ln{\left(\dot{\xi}_L/\dot{\xi_{L-1}}\right)}}
{\ln\left[L/(L-1)\right]}-1,
\label{nuest}
\end{equation}
where the dot stands for a derivative with respect to $r$. The results 
of these calculations are shown in Fig.~\ref{nu-fl}, and again they 
are consistent with the Ising value $\nu=1$.

From the measured values of the exponents $\nu$, $\eta$, and the
central charge $c$ in the thermodynamic limit, we conclude that the
isotropic--nematic transition in the fully packed limit belongs to the
Ising universality class.

\subsection{\label{sec:finite-density} Finite Fugacities}

\begin{figure}
\centering
\includegraphics[width=8cm]{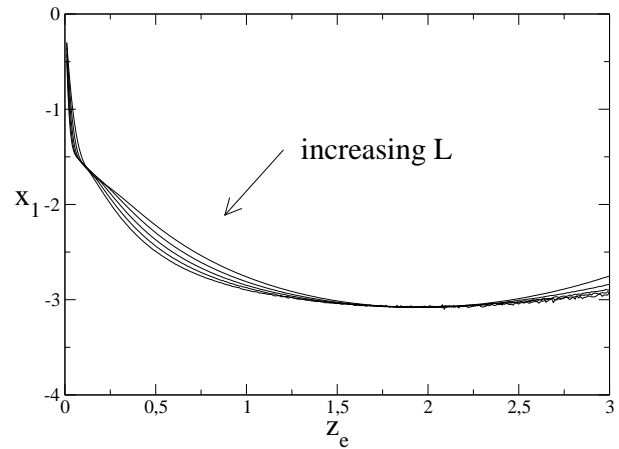}
\caption{Estimates of the exponent $x_1$ characterizing the finite
size scaling of the 
critical activity $z_{i,c}$ for fixed values of the activity $z_e$. In most 
of the range of values for the activity $z_e$, estimates obtained with
larger values of $L$ lead to smaller exponents $X_1$, but this is reversed  
for low values of $z_e$.}
\label{extrap}
\end{figure}

The procedure used to determine the critical behavior of the fully
packed limit in Sec.~\ref{sec:full-packing} is now  applied to the 
transition at finite values of $z_e$. For a fixed value of $z_e$, we
obtain estimates for the critical line [see Eq.~(\ref{prg})] 
using the correlation lengths for widths $L$ and $L+1$ 
for $L$  between 3 and 9.
The estimates for the
critical values of $z_i$ for finite width is extrapolated to infinite width 
by fitting it to the form  $z_{i,c}(L)=z_{i,c}+aL^{x_1}$. 
The estimates for the exponent $x_1$ for a range of $z_e$ are shown in 
Fig.~\ref{extrap}. As expected, a singular behavior is seen for $z_e
\to 0$, since there are no finite size corrections for $z_e=0$.  
For larger activities estimates are compatible with the Ising value 
$x_1=-3$, so we adopt this exponent for the whole critical line, 
with the exception of the singular point $z_e=0$, $z_i=1$,
for which there is no finite size dependence 
of the estimates and 
therefore $x_1=0$.

\begin{figure}
\centering
\includegraphics[width=8cm]{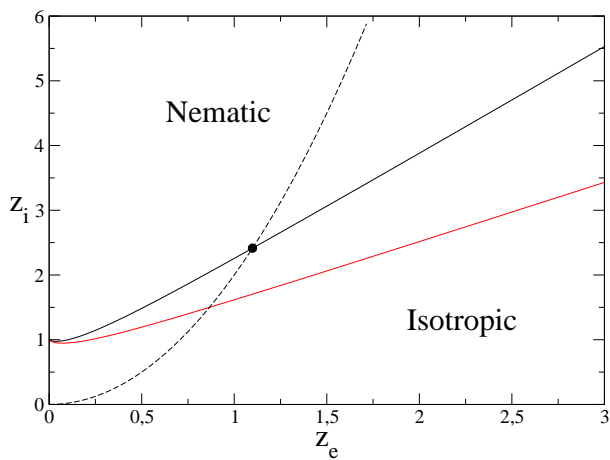}
\caption{(color on line) The phase diagram. Extrapolated transfer matrix 
estimates for the critical line (upper curve, black) and exact results 
on a four coordinated Bethe lattice (lower curve, red). On the dashed 
curve the model may be mapped on the Ising model, and the circle 
corresponds to the critical point of this model.}
\label{phdiage}
\end{figure}

Once the exponent $x_1$ is fixed, we perform two point fits to pairs of 
estimates for $z_i$. The phase diagram thus obtained is shown in
Fig.~\ref{phdiage}, where the critical
line is  the extrapolated result obtained from two point fits to the estimates 
provided by the largest widths we were able to consider, $(8,9)$ and 
$(9,10)$. The relative difference between two point fits for 
the pair of widths $(7,8)$ and $(8,9)$ and the largest widths is less 
than 0.1\% for small values of $z_e$ and decreases monotonically as 
$z_e$ grows, being of the order of $10^{-4}$ \% for $z_e=3$. This may be 
considered an estimate of the precision of the results for the critical 
line. The critical line obtained for the exact solution of the model on 
a four-coordinated Bethe lattice is also shown in Fig.~\ref{phdiage}. 
As expected, the Bethe lattice solution overestimates the nematic region. 
On the dashed line 
($z_e=2z_i^2$), the polydispersed rods model may be mapped on the Ising 
model~\cite{i06}, as discussed in Sec.~\ref{sec:intro}. The corresponding 
critical point, where this line crosses the critical line, is denoted by 
a circle.

\begin{figure}
\includegraphics[width=\columnwidth]{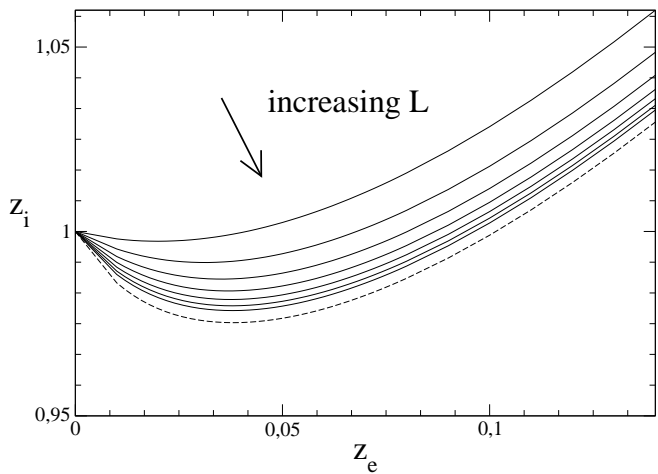}
\caption{Details of the region of small values of $z_e$ of the phase 
diagram in Fig.~\ref{phdiage}. The reentrant behavior is 
visible, as well as the estimates provided by data from distinct pairs 
of widths. For a given value of $z_e$, the estimates for the critical 
activity $z_{i,c}$ decrease as the width is increased. 
The extrapolated critical line is shown by the dashed curve.
}
\label{phdiagd}
\end{figure}

The phase diagram obtained from both transfer matrices and Bethe lattice
show a reentrant behavior at small values of $z_e$.  In Fig.~\ref{phdiagd},
we zoom in on the phase diagram for small $z_e$. Here, the 
estimates for the  critical
values of $z_i$ for all $L$ are plotted. 
The estimates follow a monotonic sequence for increasing widths, showing
a clear re-entrance in the thermodynamic limit.

\begin{figure}
\centering
\includegraphics[width=8cm]{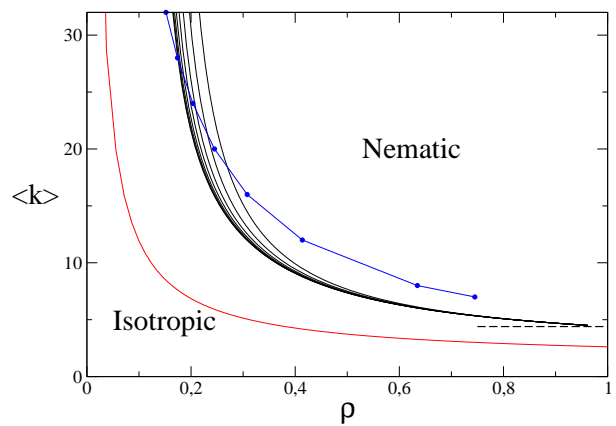}
\caption{(color on line) Phase diagram in density variables. The mean 
number of monomers per rod is shown as a function of the fraction of 
occupied lattice sites at the transition line. Results for strips of 
widths $L=3,4,5,\ldots,10$ are shown (upper curves, black), as well as 
the Bethe lattice result (lower curve, red). The dashed line corresponds 
to the estimate of $\langle k \rangle$ in the full lattice limit. 
The (blue)
dots joined by a line are data from Monte Carlo simulations of the
monodispersed case. The data are taken from Ref.~\cite{kr14}.
}
\label{phdiagdensi}
\end{figure}

Using the extrapolated data for the critical line in the parameter space 
of the activities, we may calculate the densities for the model defined 
on the strips of widths $L=3,4,\ldots,10$. The results of these 
calculations are shown in Fig.~\ref{phdiagdensi}, where the the mean 
number of monomers per rod is plotted as a function of the fraction of 
lattice sites occupied by monomers. We notice a rather strong finite 
size dependence of the results at small densities (the estimates for 
$\langle k \rangle$ are smaller for larger values of $\rho$). The 
corresponding result for the Bethe lattice solution is also shown, and
again a
larger area is occupied by the nematic phase. 
The full lattice limit 
for $\langle k \rangle$, which was estimated above, is also shown. 

We note that, contrary to what was found for monodispersed rods, a 
single transition from a disordered to a nematic phase is found for 
polydispersed rods, consistent with the results in the full lattice 
limit. Thus, in this particular polydispersed 
model the high density transition is 
absent. In
other words, while the critical $\langle k \rangle$ for the polydispersed 
model is a monotonic function of the density $\rho$, in the monodispersed 
case it displays a minimum at a value of $k$ 
between $6$ and $7$. Although the argument 
supporting a disordered phase in the full packing limit applies for all
values of $k$, the only case where the high density 
transition was studied in detail is $k=7$~\cite{krds13}.

The critical density at
the isotropic--nematic
transition for the monodispersed model was recently determined
numerically for $k$ up to $60$~\cite{kr14b}. In
Fig.~\ref{phdiagdensi},  
these data are compared with 
the phase boundary of the polydispersed model. 
For rods with widths between 5 and 6 only the polydispersed model has
a nematic phase. We notice that, in general, for a given value of 
$\langle k \rangle$ the polydispersed model orders at a smaller density, 
but apparently this is not the case at sufficiently high values of 
$\langle k \rangle$. However, the results in the low-density region may not
be sufficiently precise to confirm this conclusion. In 
this region of the parameter space the extrapolation procedure we 
adopted for the critical line may not be trustworthy, since the effective
finite size scaling exponent $x_1$ shown in Fig.~\ref{extrap} changes
rapidly 
with the activity $z_e$, due to the proximity of the singular point 
$(z_e=0,z_i=1$). Also, the numerical routines based on the power method become
very inefficient as this point is approached, due to the high degeneracy of
the leading eigenvalue at this point. Therefore, we are not able to provide 
compelling evidence of the crossing of critical lines mentioned above.

\begin{figure}
\centering
\includegraphics[width=8cm]{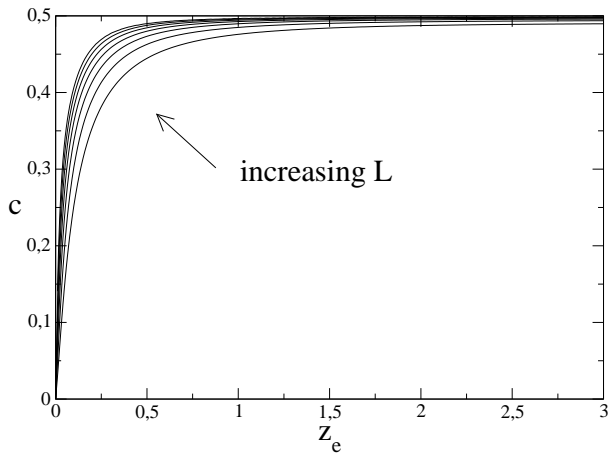}
\caption{Estimates for the central charge of the model obtained from the 
values of the free energy per site at the critical line. Larger 
estimates are obtained from data for larger widths.}
\label{c}
\end{figure}

The estimates for the central charge may now be obtained for the whole 
critical line. This is accomplished by the same procedure we used in the 
full lattice limit, through Eq. (\ref{ce}). Results for this estimate as 
a function of $z_e$ are plotted in Fig.~\ref{c}. We found estimates 
which grow with increasing widths $L$.  The lowest estimate corresponds 
to the pair of widths $L=3,4$ and the highest to $L=9,10$. At the limit 
$z_e=0$ the free energy has no finite size dependence, and therefore 
$c=0$. The estimates plotted in the graph are consistent with a step 
function $c=0$, for $z_e=0$ and $c=1/2$ for $z_e>0$, so that the Ising 
value found in the full lattice limit extends to finite non-zero values 
of $z_e$

\begin{figure}
\centering
\includegraphics[width=8cm]{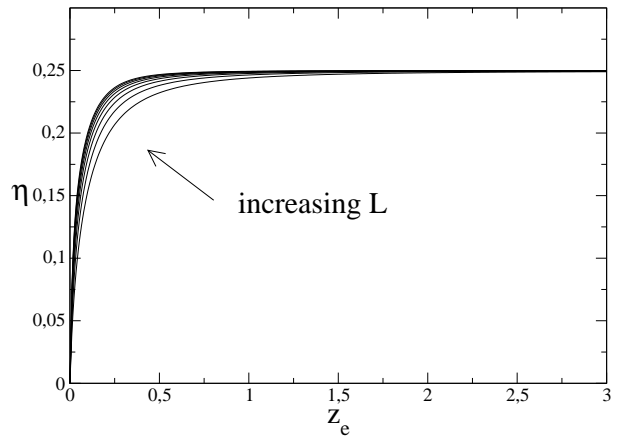}
\caption{Estimates for the exponent $\eta$ obtained from the values of 
the amplitude of the inverse correlation length at the extrapolated 
critical line. Larger estimates are obtained from data for larger 
widths.}
\label{eta}
\end{figure}

The critical exponent $\eta$ may also be estimated at the whole critical 
line using the amplitude of the correlation length, through Eq. 
(\ref{eta2}). For each of the widths, we generate an estimate for this 
exponent. The results of these calculations are shown in Fig.~\ref{eta}. 
The estimates increase with growing widths, so that the 
smallest value corresponds to $L=3$ and the largest to $L=10$. Again the 
data are consistent with an asymptotic step function behavior, with 
$\eta=0$ for $z_e=0$ and the Ising value $\eta=1/4$ for $z_e>0$.

\begin{figure}
\centering
\includegraphics[width=8cm]{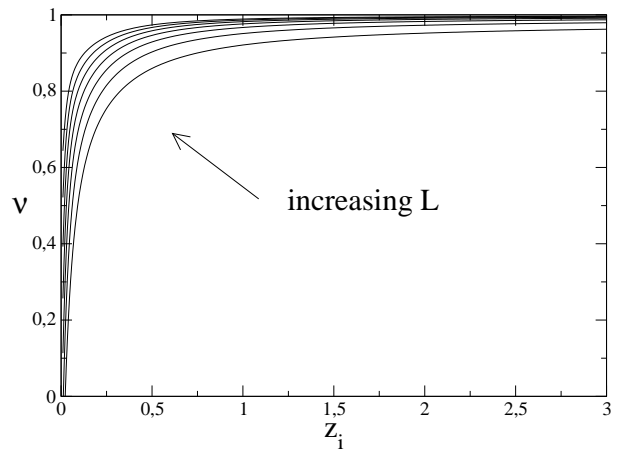}
\caption{Estimates for the exponent $\nu$ model obtained from the 
linearization of the PRG recursion relation on the critical line. Larger 
estimates are obtained from data for larger widths.}
\label{nu}
\end{figure}

Finally, estimates for the exponent $\nu$ are furnished by the 
linearization of the phenomenological renormalization group 
recursion relation at the fixed point [see Eq.~(\ref{prg})]. 
The results from such calculations, using Eq. (\ref{nuest}) 
for pairs of consecutive widths, are shown in Fig.~\ref{nu}. Again 
the results are consistent with a convergence to the asymptotic Ising 
value $\nu=1$ for $z_e>0$.

\section{Summary and Discussion}
\label{conc}

In this paper, we studied a particular model of polydispersed hard rods 
on a square lattice using transfer matrix techniques. Rods of all
sizes are allowed with the weight of a rod of length $k$ being
$z_i^{k-2} z_e^2$, where $z_i$ ($z_e$) is the activity for an internal
(endpoint) monomer. We showed that the system undergoes a single
isotropic-nematic transition as $z_i$ is increased for a fixed $z_e$.
The critical exponents and central charge obtained using finite size 
scaling are shown to be consistent with that of the two dimensional 
Ising model. 

The model studied in this paper is a variant of the model
discussed in Ref.~\cite{wkp80}, used to describe
equilibrium polymerization transitions, observed inliquid sulfur.
In this model, there are separate weights
for the endpoint monomers ($z_e$)and the internal monomers ($z_i$) 
of the polymeric chains. It may be mapped on the
magnetic $n$-vector model in the limit where the number of components
of the spin $n$ vanishes. The square of the enpoint monomer activity $z_e$ 
corresponds to the
magnetic field $h$ of the magnetic model. Therefore, the 
phase transition in this model occurs when $z_e=0$, that is, 
for infinite polymeric chains.
Also, the original model considers flexbile chains. Both these
conditions are satisfied in the equilibrium polymerization transition
in liquid sulfur. In particular, for sulfur at the 
polymerization temperature, $z_e \approx 10^{-12}$.
An experimental realization of the model 
studied in this paper would be a situation where the value of $z_i$ is finite 
and the chains are rigid. In this limit, our model predicts an
isotropic-nematic phase transition.

Contrary to what was found for monodispersed rods, the nematic phase 
remains stable up to the limit of full packing. If, in the
monodispersed system, a small density of
rods of different length are added, then the two transitions present
in the monodispersed system should persist. 
However, in the model
considered in this paper, once the density of occupied sites and the 
mean length of the rods are fixed, the polydispersity is also determined. 
It would be interesting to consider a situation where the polydispersity 
could be independently changed, in order to find out what amount of 
polydispersity is necessary to destroy the high density transition 
from the nematic to the isotropic phase. 

Our estimates for the critical parameters are consistent 
with those of the two dimensional Ising universality class, which is an
expected result if we consider the symmetry of the order parameter. This 
should change, for example, if the model is defined on the triangular or
hexagonal lattices, 
where we expect the universality class to be the that of the $q=3$ Potts model,
as was found for the isotropic-nematic transition in the 
case of monodispersed rods~\cite{flr08,flr08a}.

The transfer matrix technique used in the paper was well suited for
the particular choice of statistical weight for a rod. The number of indices
required to label a state in the transfer matrix is only $3$. 
This should be compared with the $2 k+1$ indices required to label a
state in the problem of  monodispersed rods of length $k$. This makes
it difficult to study mono-dispersed problems using transfer matrices.
However, there are some examples where transfer matrix could be
useful. One is the case of polydispersed rectangles of size $2\times
k$ with the same weight as that studied in this paper. The number of
indices to label a state is now $5$, making it suitable for transfer
matrix calculations. It would be interesting to see which of three
transitions present in the monodispersed
problem~\cite{kr14,kr14b,nkr14} persist.
Another problem is that of mixtures of hard squares and dimers which
was recently shown to have a very interesting phase diagram including
a Berezinskii-Kosterliz-Thouless transition in the fully packed
limit~\cite{ramola2014}. The number of indices now required to label a
state is only $5$ making it suitable for transfer matrix studies. The
fully packed limit is simpler since the number of states and indices are 
fewer. The same will be true for mixtures of particles with the first
$k$ nearest neighbors excluded when $k$ is
small~\cite{fernandes2007,trisha_knn}.
These are promising areas for future study.

\begin{acknowledgments}
JFS acknowledges Franciso C. Alcaraz and Marcelo S. Sarandy for helpful 
discussions and CNPq for financial support.
\end{acknowledgments}

\appendix*
\section{\label{appendix}The Gubenatis-Booth (GB) algorithm}

In the paper, the leading eigenvalues of the transfer matrix
were determined using
an algorithm that is a slightly modified version of the algorithm proposed by 
Gubenatis and Booth~\cite{gb08}. In this appendix, we briefly describe the
method. A more 
detailed discussion may be found in the original paper. 

This method is a 
variant of the power method to find the largest
eigenvalue of a real 
matrix $\mathbf{A}$.  Its simplest implementation consists of
starting with an arbitrary normalized vector $\psi$ and performing a two 
step iteration: $\phi={\mathbf A}\psi$ and $\psi=\phi/||\phi||$, where 
the normalization procedure is quite arbitrary. 
It may then be shown 
that, as long as the starting vector is not orthogonal to the (right) 
eigenvector $\psi_1$ associated with the 
largest eigenvalue $\lambda_1$ and the
eigenvalue is non-degenerate, 
the iteration 
converges to $\psi \to \psi_1/||\psi_1||$ and $||\phi|| \to \lambda_1$. 
The error, after $n$ steps, is proportional to 
$(|\lambda_2/\lambda_1|)^n$, where $\lambda_2$ is the second 
largest eigenvalue in absolute value. If the 
matrix is non-symmetric, both left and right eigenvectors need to be 
considered.

If the second largest eigenvalue is also required,
one possibility 
is to iterate
with two orthogonal vectors, so that one converges to $\psi_1$ and 
the other to $\psi_2$~\cite{w65}. 
Alternatively, the GB algorithm starts from the observation 
that any non-zero component $\alpha$ of an eigenvector belonging to 
the eigenpair $(\lambda,\psi)$ on the matrix $\mathbf{A}$
obeys
\begin{equation}
\lambda=\frac{\sum_{\beta}A_{\alpha,\beta}\psi_\beta}{\psi_\alpha}.
\end{equation}
This identity may be extended to sums of any grouping 
$R_i$ of components:
\begin{equation}
\lambda=\frac{\sum_{\alpha \in 
R_1}\sum_{\beta}A_{\alpha,\beta}\psi_\beta} {\sum_{\alpha \in 
R_1}\psi_\alpha}=\frac{\sum_{\alpha \in 
R_2}\sum_{\beta}A_{\alpha,\beta}\psi_\beta}{\sum_{\alpha \in 
R_2}\psi_\alpha}= \ldots.
\label{cons}
\end{equation}

When only the two
largest  eigenvalues are needed, the 
iterative procedure starts with two vectors 
$\psi^\prime=\sum_ia_i\psi_i$ and $\psi^{\prime\prime}=\sum_i b_i\psi_i$, 
and the behavior of the vector 
$\psi=(1-\eta)\psi^\prime+\eta\psi^{\prime\prime}$ under
iteration is considered, where $\eta$ is a number. 
Assuming that after $n$ iterations only the two dominant eigenpairs 
are significant, we have:
\begin{equation}
A^n\psi = \sum_{i=1,2}[(1-\eta)a_i+\eta b_i]\lambda_i^n\psi_i.
\end{equation}
The parameter $\eta$ is adjusted at each step to prevent the
$\psi$ from 
collapsing to the dominant eigenvector. This is accomplished by choosing 
two groupings of components $R_1$ and $R_2$ and requiring that the 
condition Eq.~(\ref{cons}) is fulfilled at each step. So, for example, 
after $n$ iterations the estimates $\kappa$ of the eigenvalue for each 
grouping are:
\begin{widetext}
\begin{eqnarray}
\kappa_1&=&\frac{[(1-\eta)a_1+\eta b_1]\lambda_1^n\sum_{\alpha \in 
R_1}\psi_{1,\alpha}+ [(1-\eta)a_2+\eta b_2]\lambda_2^n\sum_{\alpha \in 
R_1}\psi_{2,\alpha}} {[(1-\eta)a_1+\eta b_1]\lambda_1^{n-1}\sum_{\alpha 
\in R_1}\psi_{1,\alpha}+ [(1-\eta)a_2+\eta 
b_2]\lambda_2^{n-1}n\sum_{\alpha \in R_1}\psi_{2,\alpha}}, \\
\kappa_2&=&\frac{[(1-\eta)a_1+\eta b_1]\lambda_1^n\sum_{\alpha \in 
R_2}\psi_{1,\alpha}+ [(1-\eta)a_2+\eta b_2]\lambda_2^n\sum_{\alpha \in 
R_2}\psi_{2,\alpha}} {[(1-\eta)a_1+\eta b_1]\lambda_1^{n-1}\sum_{\alpha 
\in R_2}\psi_{1,\alpha}+ [(1-\eta)a_2+\eta 
b_2]\lambda_2^{n-1}n\sum_{\alpha \in R_2}\psi_{2,\alpha}},
\end{eqnarray}
\end{widetext}
Now, at each step of the iteration, one requires $\kappa_1=\kappa_2$, 
which leads to a quadratic equation for the parameter $\eta$. One 
solution $(1-\eta)a_1+b_1 \eta=0$ guides the 
process to the second 
eigenpair $\kappa_1=\kappa_2=\lambda_2$, while the other 
$(1-\eta)a_2+b_2 \eta=0$ leads it to the second eigenpair 
$\kappa_1=\kappa_2=\lambda_1$. 

Now, if the vectors at some step of the iterative
process are $\psi^\prime$ and $\psi^{\prime\prime}$, then 
we determine 
$\hat{\psi}^\prime=\mathbf{A}\psi^\prime$ and 
$\hat{\psi}^{\prime\prime}=\mathbf{A}\psi^{\prime\prime}$. The 
consistency condition will be:
\begin{equation}
\frac{(1-\eta)s_{11}+\eta s_{12}}{(1-\eta)r_{11}+\eta r_{12}}=
\frac{(1-\eta)s_{21}+\eta s_{22}}{(1-\eta)r_{21}+\eta r_{22}},
\label{cc}
\end{equation}
where the sums of components are defined as:
\begin{eqnarray}
s_{i,1}&=&\sum_{\alpha \in R_i}\hat{\psi}^\prime_\alpha, \\
s_{i,2}&=&\sum_{\alpha \in R_i}\hat{\psi}^{\prime\prime}_\alpha, \\
r_{i,1}&=&\sum_{\alpha \in R_i}\psi^\prime_{\alpha}, \\
r_{i,2}&=&\sum_{\alpha \in R_i}\psi^{\prime\prime}_{\alpha}.
\end{eqnarray}

The consistency condition Eq.~(\ref{cc}) 
leads to a quadratic equation for the 
parameter $\eta$:
\begin{equation}
a_2 \eta^2+a_1 \eta+a_0=0,
\label{etaeq}
\end{equation}
where the coefficients are given in terms of the sums:
\begin{eqnarray}
a_0&=&r_{12}s_{11}-r_{11}s_{12}, \\
a_1&=&2(r_{11}s_{12}-r_{12}s_{11})+r_{22}s_{11}+r_{12}s_{21}
\nonumber \\ 
&& - r_{21}s_{12}
-r_{11}s_{22}, \\
a_2&=&r_{12}s_{11}+r_{22}s_{21}+r_{21}s_{12}+r_{11}s_{22}
\nonumber \\ 
&&
-r_{11}s_{12}-
r_{21}s_{22}-r_{22}s_{11}-r_{12}s_{21}.
\end{eqnarray}

Further refinements of the algorithm that results in
faster convergence may be found in the original article~\cite{gb08}.

We summarize the steps of the algorithm.
\begin{enumerate}

\item 

Define small convergence parameters $\epsilon_1$ and $\epsilon_2$. In our 
double precision calculations, we have chosen 
$\epsilon_1=\epsilon_2=10^{-13}$.

\item 

Make initial guesses
for vectors $\psi^\prime$ and $\psi^{\prime\prime}$, 
and the groupings of components 
$R_1$ and $R_2$. If the matrix 
$\mathbf{A}$ is of size $N \times N$, we have chosen 
$R_1=(1,2,\ldots,[N/2])$ and $R_2=[N/2]+1,[N/2]+2,\ldots,N)$.

\item 

Normalize: $\psi^\prime \leftarrow \psi^\prime/||\psi^\prime||$, 
$\psi^{\prime\prime} \leftarrow 
\psi^{\prime\prime}/||\psi^{\prime\prime}||$.

\item 

Obtain $\hat{\psi^\prime}=\mathbf{A}\psi^\prime$ and 
$\hat{\psi^{\prime\prime}}=\mathbf{A}\psi^{\prime\prime}$.

\item 

Find the two values of $\eta$ for which consistency condition
[see Eq.~(\ref{etaeq})] is satisfied. 
If the roots are complex, update 
vectors ($\psi^\prime \leftarrow \hat{\psi^\prime}$ and 
$\psi^{\prime\prime} \leftarrow \hat{\psi^{\prime\prime}}$ and return to 
step 3.

\item 

Find estimates of eigenvalues associated to the two real values of 
$\eta$, using, for example, the grouping $R_1$: \begin{equation} 
\lambda=\frac{(1-\eta)s_{11}+\eta s_{21}}{(1-\eta)r_{11}+\eta r_{21}}. 
\end{equation} Define $\eta_1$ to be the solution associated to the 
largest eigenvalue $\lambda_1$ and $\eta_2$ related to the second 
largest eigenvalue $\lambda_2$.

\item 

Check for convergence: If $|\eta_1|>\epsilon_1$ or 
$|1-\eta_2|>\epsilon_2$ update vectors $\psi^\prime \leftarrow 
(1-\eta_1)\hat{\psi^\prime}+\eta_1 \hat{\psi^{\prime\prime}}$, 
$\psi^{\prime\prime} \leftarrow (1-\eta_2) \hat{\psi^\prime} +\eta_2 
\hat{\psi^{\prime\prime}}$ and return to step 3.

\item 

Terminate.

\end{enumerate}


%

\end{document}